\documentstyle[12pt]{article}

\parskip=\medskipamount
\begin{document}

\begin{titlepage}

\begin{center}
{ \large \bf Intensity-dependent pion-nucleon coupling and the
Wr\' oblewski  relation } \\
\vspace{2.cm}
{ \normalsize \sc  M. Martinis and V. Mikuta-Martinis\footnote
{e-mail address: martinis@thphys.irb.hr \\
e-mail address: mikutama@thphys.irb.hr}} \\
\vspace{0.5cm}
Department of  Physics, \\
Theory Division, \\
Rudjer Bo\v skovi\' c Institute, P.O.B. 1016, \\
10000 Zagreb,CROATIA \\
\vspace{2cm}
{ \large \bf Abstract}
\end{center}
\vspace{0.5cm}
\baselineskip=24pt
We propose an intensity-dependent pion-nucleon coupling Hamiltonian
within a unitary multiparticle-production  model of the
Auerbach-Avin-Blankenbecler-Sugar (AABS) type in which
the  pion field is represented
by the thermal-density matrix. Using this Hamiltonian, we explain
the appearance of the negative-binomial (NB) distribution for pions and
the well-known empirical relation, the so-called Wr\' oblewski relation,
in which the  dispersion $D$ of the pion-multiplicity distribution
is linearly related to the average multiplicity $<n>$ :
$D = A<n> + B $, with the coefficient $A $ related to the vacuum
energy of the pion field and $< 1$ .
 The Hamiltonian of our  model is expressed
linearly in terms of the generators of the $SU(1,1)$ group.
We also find the generating function for the pion field.
All higher-order thermal moments can be calculated from this function.
At $T = 0$ it reduces to  the generating function of the NB distribution.

\end{titlepage}

\setcounter{page}{2}
\newpage
\baselineskip=24pt
\vspace{1.5cm}
\section{Introduction}

During the last years a considerable amount of experimental information
has been accumulated on multiplicity distributions of charged
particles produced in $pp$ and $p\bar{p}$ collisions in the centre-
of-mass energy range from $10 GeV$  to $1800 GeV$ .
Measurements in the regime of several hundred $GeV$ [1] have shown
the violation of the Koba-Nielsen-Olesen (KNO)
scaling [2], which was previously observed in the
 ISR c.m.energy range from 11 to 63 $GeV$  [3].
The violation of the KNO scaling is characterized by an
enhancement of high-multiplicity events leading to a
broadening of the multiplicity distribution with energy.

The shape of the multiplicity distribution may be described
either by its \\
C moments, $C_{q} = \langle n^{q}\rangle /
\langle n\rangle ^{q}$, or by  its central moments (higher-order dispersions),
$D_{q} = \langle (n - \langle n\rangle )^{q}
\rangle ^{1/q},  q = 2,3 \ldots $.
The exact KNO scaling implies that all $C_{q}$
moments are energy independent. Only at energies below $100 GeV$
do the $C$ moments appear to be energy independent.
It can  also been shown [4] that the KNO scaling  leads to a
generalized Wr\' oblewski relation  [5]
\begin{equation}
D_{q} = A_{q}\langle n\rangle  - B_{q},
\end{equation}
with the energy-independent coefficients $A_{q}$ and $B_{q}$.
The $pp$ and $p\bar{p}$ inelastic data below $100 GeV$  also
show the linear dependence of the dispersion on
the average number of charged particles, but with the coefficients
$A_{q}$ and $B_{q}$ that are approximately equal within errors.

The fact that the dispersion of the multiplicity distribution
grows linearly with $\langle n\rangle $ implies that the
elementary Poisson distribution resulting from the independent
emission of particles is ruled out.

The total multiplicity distribution $P_{n}$ of charged particles
for a wide range of energies $(22 - 900 GeV)$
is found to be well described by a negative-binomial (NB)
distribution [1,6] that belongs
to a large class of compound Poisson distributions [7].
The compound Poisson distribution is
fully determined by its generating function $G(z)$ of the form
\begin{equation}
G(z) =  e^{\bar{N}[g(z) - 1]} = \sum z^{n}P_{n},
\end{equation}
where $\bar{N}$ is the average number of $N$ independently produced
clusters or clans (Poisson distributed) [8]. Each of these
subsequently decays according to the probability distribution
corresponding to the generating function $g(z)$.
The NB distribution is obtained by choosing $g(z) = ln(1 - pz)/
ln(1 - p)$, with $p = (1 - \langle n\rangle /k)^{-1}$.
It is a two-step process [8] with two free parameters:
the average number of charged particles  $\langle n\rangle $
and the parameter $k$ which affects the shape (width) of the
distribution. The parameter $k$ is also related to the
dispersion $D = D_{2}$ by the relation
\begin{equation}
(\frac{D}{\langle n\rangle })^{2}  = \frac{1}{k} + \frac{1}{\langle n\rangle },
\end{equation}
so that the observed broadening of the normalized multiplicity
distribution with increasing energy implies a decrease of the
parameter $k$ with energy. The KNO scaling requires constant $k$.

Although the NB distribution
gives information on the structure of correlation functions in
multiparticle production,  the question still remains
whether its clan-structure interpretation  is
simply a new parametrization of the data or  has a deeper physical
insight [9]. Measurements  of multiplicity distributions in $p\bar{p}$
collisions at $TeV$ energies [10] have recently shown that their shape
is clearly different from that of the NB distribution.
The distributions display the so-called medium-multiplicity
"shoulder", with a shape qualitatively similar to that
of the UA5 $900 GeV$  and UA1 distributions [11]. A satisfactory
explanation of this effect is still lacking [12].

In this paper we propose another approach to multiplicity
distributions based on a unitary eikonal model with a
pion-field thermal-density operator given in terms of an effective
intensity-dependent pion-nucleon coupling Hamiltonian. We assume that
the system of produced hadronic matter (pions) is in thermal
equilibrium at the temperature $T$ immediately after the collision.

The paper is organized as follows.
In Sect.2  we explain the basic ideas of our unitary eikonal model
with a  pion-field thermal-density operator.
A discussion of the Wr\' oblewski relation is presented in Sect. 3.
Finally, in Sect. 4 we draw conclusions and make remarks on the possible
extension of the model to include  two-pion correlations
in the effective pion-nucleon Hamiltonian.

\newpage
\section{Description of the model}

At present accelerator energies the number of secondary
particles (mostly pions) produced in hadron-hadron collisions
is large enough, so that the statistical approach to particle
production becomes reasonable . Most of the properties of
pions produced in  high-energy hadron-hadron collisions
can be expressed simply in terms of a pion-field density
operator. We neglect difficulties associated with isospin
and only consider the production of  isoscalar "pions".
We expect that in high-energy collisions most of the pions
are produced in the central region. In this region the energy - momentum
conservation has a minor effect if the transverse momenta of the
pions are limited by the dynamics and their rapidities restricted
to the range $\mid y\mid < Y$, where $Y= ln(s/m^{2})$ is the
relative rapidity of the colliding particles.

{\bf A. The AABS model}

A long time ago Auerbach, Avin, Blankenbecler, and Sugar [13] presented
a class of models (AABS models)
for which the scattering operator satisfied the  exact s-channel
unitarity at high energies. In these models, the incident hadrons propagate
through the interaction region, without making significant
changes in their longitudinal momenta (leading-particle effect).
Only the part $W = K\sqrt{s}$ of the total c.m. energy $\sqrt{s}$  in
every concrete event is avaliable for particle production, where
$K$ is the inelasticity: $0\le K \le 1$ .

In the AABS type of models, the scattering operator $\hat{S}$ is
diagonal in the rapidity difference $Y$  and in the relative impact
parameter $\vec{B}$ of the two incident hadrons.
The initial-state vector for the pion field
is  $\hat{S}(Y,\vec{B})\mid 0 >$, where the vacuum state $\mid 0 >$
for pions is in fact a state containing two incident hadrons.

The $n$-pion production amplitude for $n\ge 1$ is  given  by
\begin{equation}
iT_{n}(Y,\vec{B}; k_{1}\ldots k_{n}) = 2s\langle k_{1}\ldots k_{n}
\mid \hat{S}(Y,\vec{B})\mid 0\rangle .
\end{equation}
We write the square of the $n$-pion production amplitude in the form
\begin{equation}
\mid T_{n}(Y,\vec{B}; k_{1}\ldots k_{n})\mid ^{2} =
4s^{2}Tr\{\rho (Y,\vec{B})\mid k_{1}\ldots k_{n}\rangle \langle k_{1}
\ldots k_{n}\mid \},
\end{equation}
where the pion-density operator $\rho (Y,\vec{B})$ is defined as
\begin{equation}
\rho (Y,\vec{B}) = \hat{S}(Y,\vec{B})\mid 0\rangle \langle 0\mid
\hat{S}^{\dag }(Y,\vec{B}).
\end{equation}
The square of the elastic scattering amplitude is then  the
matrix element of $\rho (Y,\vec{B})$ between the states with no
pions, i.e., $\langle 0\mid \rho (Y,\vec{B})\mid 0\rangle $.
In terms of the pion-number operator
\begin{equation}
\hat{N} = \sum_{k} a^{\dag }_{k}a_{k} = \sum_{k} \hat{N}_{k} , \,\,
k \equiv (\omega _{k},\vec{k}),
\end{equation}
the square of the $S$-matrix element when no pions are emitted can
also be written in the form
\begin{equation}
\mid \langle 0\mid \hat{S}(Y,\vec{B})\mid 0\rangle \mid ^{2} =
Tr\{ \rho (Y,\vec{B}) : e^{-\hat{N}} :\} = e^{- \Omega (Y,\vec{B})}.
\end{equation}
Here $: \, :$ indicates the operation of normal ordering and
$\Omega (Y,\vec{B})$ is the usual eikonal function (or the
opacity function) of the geometrical model [14]. The connection
with the inelastic cross section and the exclusive cross section
for production of $n$ pions is then
\begin{equation}
\sigma _{inel}(Y,\vec{B}) = 1 - e^{- \Omega (Y,\vec{B})},
\end{equation}
and for $n\geq 1$, it is
\begin{equation}
\sigma _{n}(Y,\vec{B}) = Tr\{ \rho (Y,\vec{B}) : \frac{\hat{N}^{n}}{n!}
e^{- \hat{N}} :\}.
\end{equation}
In terms of a normalized pion-multiplicity distribution at each impact
parameter, $P_{n}(Y,\vec{B}) = \sigma _{n}
(Y,\vec{B})/ \sigma _{inel}(Y,\vec{B})$, the observed complete
multiplicity distribution $P_{n}(Y)$ is obtained by summing $P_{n}(Y,\vec{B})$
over all impact parameters $\vec{B}$ with the weight function
$Q(Y,\vec{B}) = \sigma_{inel}(Y,\vec{B}) / \sigma_{inel}(Y)$, i.e.,
\begin{equation}
P_{n}(Y) = \int d^{2}B Q(Y,\vec{B}) P_{n}(Y,\vec{B}).
\end{equation}

The first-order moment of $P_{n}(Y)$ is the average  multiplicity
\begin{equation}
\langle n\rangle = \sum n P_{n}(Y) = \int d^{2}B Q(Y,\vec{B})
\bar{n}(Y,\vec{B}).
\end{equation}
The higher-order moments of $P_{n}(Y)$ give information on the
dynamical fluctuations from $\langle n\rangle $ and also on the
multiparticle correlations. All these higher-order moments can be
obtained from the pion-generating function
\begin{equation}
G(z) = \sum z^{n}P_{n}(Y) = \int d^{2}B Q(Y,\vec{B})G(Y,\vec{B};z),
\end{equation}
by differentiation where
\begin{equation}
G(Y,\vec{B};z) = Tr\{\rho (Y,\vec{B})z^{\hat{N}}\}
\end{equation}
is the pion-generating function in $B$-space .
Thus the normalized factorial moments $F_{q}$ are
\begin{equation}
F_{q} = \frac{\langle n(n-1)\ldots (n-q+1)\rangle}{\langle n\rangle ^{q}} =
\langle n\rangle ^{- q}\frac{d^{q}G(1)}{dz^{q}}
\end{equation}
and the normalized cumulant moments $K_{q}$ are
\begin{equation}
K_{q} = \langle n\rangle ^{- q}\frac{d^{q}lnG(1)}{dz^{q}}.
\end{equation}
These moments are related to each other by the formula
\begin{equation}
F_{q} = \sum_{l=0}^{q-1} {q-1 \choose l}K_{q-l}F_{l}.
\end{equation}
For the Poisson distribution, all the normalized factorial moments are
identically equal to $1$ and all  cumulants vanish for $q > 1$.

We are concerned here mostly with the $q =2$ moments, which are directly
related to the dispersion $D$:
\begin{equation}
F_{2} = K_{2} + 1 = (\frac{D}{\langle n\rangle })^{2} +
1 - \frac{1}{\langle n\rangle }.
\end{equation}

{\bf B. Thermal-density operator for the pion field}

The operator $\mid 0\rangle \langle 0\mid $ appearing in the definition
of $\rho (Y,\vec{B})$ represents the density operator $\rho (vac)$ for
the pion-field vacuum state. It can also be considered as describing the pion
system in thermal equilibrium at the temperature $T = 0$.
The density operator for a
pion field in thermal equilibrium at the temperature $T$ is then
\begin{equation}
\rho _{T} = \frac{1}{Z} e^{- \beta H_{0}} , \,\, \beta = \frac{1}{k_{B}T},
\end{equation}
where
\begin{eqnarray}
H_{0} & = & \sum_{k} \omega _{k}(a^{\dag }_{k}a_{k} + \lambda ), \\
  lnZ & = & - \beta \lambda \sum_{k} \omega _{k} -
\sum_{k} ln(1 - e^{-\beta \omega _{k}}). \nonumber
\end{eqnarray}
The quantity $\lambda \sum_{k}\omega _{k}$ represents the lowest possible
energy of the pion system at the temperature $T = 0$.  The " zero-point
energy " corresponds to $\lambda = \frac{1}{2}$.
If the pion energies $\omega _{k} = \sqrt{\vec{k}^{2} + m_{\pi }^{2}}$ are
closely spaced, the summation over $k$ is replaced by an integral:
$\sum_{k} \to \int d^{3}k/2\omega _{k}$. \\
Note that $\rho (vac) = \rho _{T=0}$.
The mean number of thermal (chaotic) pions is
\begin{eqnarray}
\bar{n}_{T} & = & \sum_{k} \frac{1}{e^{\beta \omega _{k}} - 1}  \\  \nonumber
& = & \sum_{k}\bar{n}_{Tk}.
\end{eqnarray}

The transformed thermal-density operator $\rho _{T}(Y,\vec{B})$ is now
\begin{eqnarray}
\rho _{T}(Y,\vec{B}) & = & \hat{S}(Y,\vec{B}) \rho _{T}
\hat{S^{\dag}}(Y,\vec{B}) \\  \nonumber
                     & = & \frac{1}{Z}e^{- \beta H(Y,\vec{B})},
\end{eqnarray}
where
\begin{equation}
H(Y,\vec{B}) = \hat{S}(Y,\vec{B})H_{0}\hat{S^{\dag}}(Y,\vec{B})
\end{equation}
is regarded as an effective Hamiltonian describing the pion system
in the presence of two leading particles (nucleons).
Taking into account an old observation of Golab-Meyer and Ruijgrok [15]
that the Wr\' oblewski relation can be satisfied for all energies if
the square of the pion-nucleon coupling constant increases linearly
with the mean number of pions $\langle n\rangle $, we propose the
following form of the effective pion-nucleon
intensity-dependent coupling Hamiltonian:
\begin{eqnarray}
H(Y,\vec{B}) & = & \sum_{k} [ \epsilon _{k}(Y,\vec{B})(N_{k} + \lambda ) +
g_{k}(Y,\vec{B})( a_{k}\sqrt{N_{k} + 2\lambda  - 1} + h.c.) ] \\
             & = & \sum_{k} H_{k}(Y,\vec{B}),  \nonumber
\end{eqnarray}
where $\epsilon _{k}^{2}(Y,\vec{B}) = \omega _{k}^{2} + 4g_{k}^{2}(Y,\vec{B})$.
The interaction part of the Hamiltonian $H_{k}$ for the k mode is no longer
linear in the pion-field variables $a_{k}$ and represents an
intensity-dependent coupling [16]. It is also easy to see that the operators
\begin{eqnarray}
K_{0}(k) & = & N_{k} + \lambda  , \\  \nonumber
K_{-}(k) & = & a_{k}\sqrt{N_{k} + 2\lambda - 1} ,  \\
K_{+}(k) & = & \sqrt{N_{k} + 2\lambda - 1}\,a_{k}^{\dag }  \nonumber
\end{eqnarray}
form the standard Holstein-Primakoff [17] realizations of the $su(1,1)$ Lie
algebra, the Casimir operator of which is
\begin{equation}
\hat{C}_{k} = K_{0}^{2}(k) - \frac{1}{2}[K_{+}(k)K_{-}(k) + K_{-}(k)K_{+}(k)] =
\lambda (\lambda  - 1)\hat{I}_{k}.
\end{equation}
The Hamiltonian  $H_{k}(Y,\vec{B})\equiv  H_{k}$ is thus a linear
combination of the generators of the  $SU(1,1)$ group:
\begin{equation}
H_{k} = \epsilon _{k} K_{0}(k) + g_{k}[K_{+}(k) + K_{-}(k)].
\end{equation}
The corresponding S-matrix which diagonalizes the Hamiltonian $H(Y,\vec{B})$
is
\begin{equation}
\hat{S}(Y,\vec{B}) = \prod_{k} \hat{S}_{k}(Y,\vec{B}),
\end{equation}
where
\begin{equation}
\hat{S}_{k}(Y,\vec{B}) = exp\{ - \theta _{k}(Y,\vec{B})[K_{+}(k) -
K_{-}(k)]\},
\end{equation}
with
\begin{equation}
th\,\theta _{k}(Y,\vec{B}) = \frac{2g_{k}(Y,\vec{B})}{\epsilon
_{k}(Y,\vec{B})}.
\end{equation}
Since the dependence on the variables $Y,\vec{B}$ is contained only
in the hyperbolic angle $\theta _{k}(Y,\vec{B})$, we shall from now on
assume this dependence whenever we write $\theta _{k}$.

It is  easy to see that the initial-state vector for the pion field,
$\hat{S}(Y,\vec{B})\mid 0\rangle $, factorizes
\begin{equation}
\hat{S}(Y,\vec{B})\mid 0\rangle  = \prod _{k}(\hat{S}_{k}(Y,\vec{B})
\mid 0_{k}\rangle ),
\end{equation}
with
\begin{eqnarray}
\hat{S}_{k}(Y,\vec{B})\mid 0_{k}\rangle & = & (1 - th^{2}\theta _{k})^{\lambda
}
\sum _{n_{k}}(- th\theta _{k})^{n_{k}}\big(\frac{\Gamma (n_{k} + 2\lambda )}
{n_{k}!\Gamma (2\lambda )}\big)^{1/2}\mid n_{k}\rangle  \\
& = & \mid \theta _{k}\rangle .  \nonumber
\end{eqnarray}
where $\mid n_{k}\rangle  = (n_{k}!)^{- 1/2}(a^{\dag }_{k})^{n_{k}}
\mid 0_{k}\rangle $.
In the same way we find that the pion thermal-density operator
$\rho _{T}(Y,\vec{B})$ is also factorized as
\begin{equation}
\rho _{T}(Y,\vec{B}) = \prod _{k}\rho _{T}(\theta _{k}),
\end{equation}
with
\begin{equation}
\rho _{T}(\theta _{k}) = \frac{1}{Z_{k}}\sum _{n_{k}} e^{- \beta \omega _{k}
(n_{k} + \lambda )}\mid n_{k},\theta _{k}\rangle \langle n_{k},\theta _{k}\mid
,
\end{equation}
where $\mid n_{k},\theta _{k}\rangle  = \hat{S}_{k}(Y,\vec{B})\mid n_{k}\rangle
$.
Note that $\mid n_{k},\theta _{k}\rangle $ form a  complete orthonormal
set of eigenvectors of the $k$-mode
Hamiltonian $H_{k}$, i.e.,
\begin{equation}
H_{k}\mid n_{k},\theta _{k}\rangle   = \omega _{k}(n_{k} + \lambda )
\mid n_{k},\theta _{k}\rangle .
\end{equation}

\newpage
\section{Pion-generating function and its moments }

The average multiplicity $\bar{n}_{T}(Y,\vec{B})$, the dispersion
$d_{T}^{2}(Y,\vec{B})$, and all  higher-order moments
\begin{equation}
\bar{n^{q}_{T}}(Y,\vec{B}) = Tr\{\rho _{T}(Y,\vec{B})\hat{N}^{q}\},\,
q = 1,2,\ldots
\end{equation}
at the temperature $T$ in $B$ space can be obtained
from the pion-generating function
\begin{equation}
G_{T}(Y,\vec{B};z) = \prod _{k}G_{T}(\theta _{k};z)
\end{equation}
by differentiation, where
\begin{equation}
G_{T}(\theta _{k};z) = Tr\{\rho _{T}(\theta _{k})z^{\hat{N}_{k}}\}.
\end{equation}

After performing a certain amount of straightforward
algebraic manipulations, we arrive [18]
at the following expression for the pion-generating function
$G_{T}(\theta _{k};z)$:
\begin{equation}
G_{T}(\theta _{k};z) = G_{0}(\theta _{k};z)(1 - e^{- \beta \omega _{k}})
2^{2\lambda - 1}R_{k}^{-1}(1 + y_{k}  + R_{k})^{1 - 2\lambda },
\end{equation}
where
\begin{eqnarray}
R_{k} & = & \sqrt{1 - 2x_{k}y_{k} + y_{k}^{2}},  \\   \nonumber
x_{k} & = & \frac{z + (1-z)^{2}sh^{2}(\theta _{k})ch^{2}(\theta _{k})}
{z - (1-z)^{2}sh^{2}(\theta _{k})ch^{2}(\theta _{k})},  \\
y_{k} & = & e^{-\beta \omega _{k}}\frac{z - (1-z)sh^{2}(\theta _{k})}
{1 + (1-z)^{2}sh^{2}(\theta _{k})}.  \nonumber
\end{eqnarray}
and $G_{0}(\theta _{k};z)$ denotes the pion-generating function
at the temperature $T = 0$:
\begin{equation}
G_{0}(\theta _{k};z) = [ 1 + (1-z)sh^{2}(\theta _{k})]^{- 2\lambda }.
\end{equation}
We observe that $G_{0}$ is exactly the generating function of the NB
distribution with a constant shape parameter  $2\lambda $, and the
average number of k-mode pions is equal to
\begin{equation}
\bar{n}(\theta _{k}) = 2\lambda sh^{2}(\theta _{k}).
\end{equation}
The vacuum value of the $k$-mode thermal-density operator
$\rho _{T}(\theta _{k})$ is used to obtain the
$k$-mode thermal eikonal function $\Omega _{T}(\theta _{k})$
\begin{eqnarray}
\langle 0_{k}\mid \rho _{T}(\theta _{k})\mid 0_{k}\rangle  & = &
e^{-\Omega _{T}(\theta _{k})}  \\  \nonumber
& = & (1 - e^{- \beta \omega _{k}})G_{0}(\theta _{k};e^{- \beta \omega _{k}}).
\end{eqnarray}
The total eikonal function is
$\Omega _{T}(Y,\vec{B}) = \sum_{k}\Omega _{T}(\theta _{k}) $.

For the $k$-mode pion field in $B$ space at the temperature $T$, we
find the following average number and the dispersion:
\begin{eqnarray}
\bar{n}_{T}(\theta _{k}) & = & \bar{n}(\theta _{k}) + \bar{n}_{Tk} +
\frac{1}{\lambda }\bar{n}(\theta _{k})\bar{n}_{Tk},  \\  \nonumber
d_{T}^{2}(\theta _{k}) & = & d_{Tk}^{2} + d^{2}(\theta _{k})
[ 1 + \frac{2\lambda - 3}{\lambda }\bar{n}_{Tk} + \frac{4}{\lambda }
\bar{n}_{Tk}^{2}],
\end{eqnarray}
where
\begin{eqnarray}
d_{Tk}^{2} & = & \bar{n}_{Tk}^{2} + \bar{n}_{Tk},  \nonumber  \\
d^{2}(\theta _{k}) & = & \frac{1}{2\lambda }\bar{n}^{2}(\theta _{k}) +
\bar{n}(\theta _{k}).
\end{eqnarray}

Two limiting cases are of interest namely, $T = 0$ and $T \to\infty $.
For the $T = 0$ case, we find
\begin{equation}
\frac{d^{2}(\theta _{k})}{\bar{n}^{2}(\theta _{k})} =
\frac{1}{2\lambda } + \frac{1}{\bar{n}(\theta _{k})},
\end{equation}
as it is to be expected from the  NB distribution ( Eq. 3).
However,  our interpretation of this result is quite different.
In our case, the parameter $\lambda $ is connected with
the vacuum energy of the pion field in Eq. 20, and has nothing to do
with either the number of pion sources or the number of clans.
It is a real and positive constant labeling the positive discrete class of
a unitary irreducible representation of $SU(1,1)$, which is a
dynamical symmetry group of our system Hamiltonian. It is important
to observe that pions in the $k$ mode are distributed according to
the NB distribution with a constant shape parameter $2\lambda $.
The Wr\' oblewski relation
\begin{equation}
d(\theta _{k}) = A\bar{n}(\theta _{k}) + B
\end{equation}
is obtained with energy-independent coefficients $A = (2\lambda )^{- 1/2}$
and $B = (\lambda /2)^{1/2}$. If $\lambda  > 1/2$, we have $A < 1$.

 The contribution from all the $k$ modes gives
\begin{equation}
\frac{d^{2}(Y,\vec{B})}{\bar{n}^{2}(Y,\vec{B})} = \frac{1}{2\lambda }
\sum_{k}p^{2}(\theta _{k})  +  \frac{1}{\bar{n}(Y,\vec{B})},
\end{equation}
where $p(\theta _{k}) = \bar{n}(\theta _{k})/\bar{n}(Y,\vec{B})$. In
this case, the coefficient $A$ in the Wr\' oblewski relation becomes
energy dependent and is of the form
\begin{equation}
A(Y,\vec{B}) = \big[ \frac{1}{2\lambda }\sum_{k}p^{2}(\theta _{k})\big]^{1/2}.
\end{equation}
Since $\sum_{k}p(\theta _{k}) = 1$ and all $p(\theta _{k})$ are positive
functions of $\theta _{k}$, the sum $\sum_{k}p^{2}(\theta _{k})$ is always
smaller than one. Therefore, $A(Y,\vec{B}) < 1$ if $\lambda  > 1/2$.

Finally, the summation over all impact parameters gives
\begin{equation}
\big(\frac{D}{\langle n\rangle }\big)^{2} = \int d^{2}B Q(Y,\vec{B})
[(A^{2}(Y,\vec{B}) + 1)\big(\frac{\bar{n}(Y,\vec{B})}
{\langle n\rangle }\big)^{2} - 1] + \frac{1}{\langle n\rangle }.
\end{equation}
This expression, when combined with our preceding analysis suggests that
the coefficient $A$ in the Wr\' oblewski relation  should be energy
dependent and smaller than one.

For the temperature $T$ going to infinity we obtain
\begin{eqnarray}
\frac{d_{T}^{2}(\theta _{k})}{\bar{n}_{Tk}^{2}(\theta _{k})}
\Big|_{T\to\infty} & = & 2 - (1 + \frac{\bar{n}(\theta _{k})}
{\lambda })^{- 2}   \\   \nonumber
& = &  1 + th^{2}(2\theta _{k}).
\end{eqnarray}
This result shows that at very high temperature the distribution of
pions  will become chaotic if $\theta _{k}$ is very small. This will
happen when $\omega _{k} \gg g_{k}(Y,\vec{B})$.

\newpage
\section{Conclusions}

In this paper we have proposed an intensity-dependent pion-nucleon
coupling Hamiltonian with $SU(1,1)$  dynamical symmetry, within
a multiparticle-production model of the AABS type
in which the $k$ mode pion field is represented by the thermal-density
operator. We have shown that this Hamiltonian
explains in a natural way the appearance of the NB
multiplicity distribution for pions in impact-parameter space.The shape
parameter of the NB distribution is related to the vacuum energy
of the pion field at the temperature $T = 0$.

The Wr\' oblewski relation is obtained with the coefficient $A$
that is energy dependent  and smaller than one if the vacuum energy
of the pion field is larger than "zero-point energy"
corresponding to $\lambda = 1/2$.

For $T \neq 0$, we have found a pion-generating function that may
be used for obtaining all  higher-order  moments of the pion field.

In our model, the $k$ modes of the pion field are statistically independent
and are described by the factorized thermal-density operator. Correlations
between different $k$ modes are absent and, at this stage, our model
cannot describe the emission of resonances. However, this can be remedied
by adding  a mode-mode interacting part to the total Hamiltonian
$H(Y,\vec{B})$, e.g.,
$\sum_{k,k'}[A_{k,k'}(Y,\vec{B})a_{k}a_{k'} + h.c. ]$ [19].  We hope
to treat this case elsewhere.

\vspace{1cm}

{ \large \bf Acknowledgment }

This work was supported by the Ministry of Science of the Republic
of Croatia under Contract No.1-03-212.

\end{document}